\documentclass[%
final,
 aip,
 jcp,
 amsmath,amssymb,
 preprint,%
floatfix,
]{revtex4-2}

\usepackage{graphicx}
\usepackage{dcolumn}
\usepackage{bm}
\usepackage{mathdots} 
\usepackage[mathlines]{lineno}
\usepackage{csquotes}
\usepackage{upgreek}
\usepackage{subcaption}
\usepackage{chemmacros}
\usepackage{algorithmic}
\usepackage{algorithm}
\usepackage{multirow}
\usepackage{cleveref}
\DeclareSIUnit[per-mode=reciprocal]{\wn}{\cm\tothe{-1}}
\begin{document}
\title
{
The Conundrum of Diffuse Basis Sets: A Blessing for Accuracy yet a Curse for Sparsity
}

\author{Henryk Laqua}
\affiliation{
Kenneth S. Pitzer Center for Theoretical Chemistry, Department of Chemistry, University of California, Berkeley, CA 94720, USA.
}%
\author{Linus Bjarne Dittmer}
\affiliation{
Kenneth S. Pitzer Center for Theoretical Chemistry, Department of Chemistry, University of California, Berkeley, CA 94720, USA.
}%
\affiliation{ 
Interdisciplinary Center for Scientific Computing, Ruprecht-Karls University, Im Neuenheimer Feld 205, 69120 Heidelberg, Germany
}%
\author{Martin Head-Gordon}
\email{m\_headgordon@berkeley.edu}
\affiliation{
Kenneth S. Pitzer Center for Theoretical Chemistry, Department of Chemistry, University of California, Berkeley, CA 94720, USA.
}%
\date{\today}
\begin{abstract}
Diffuse atomic orbital basis sets have proven to be essential to obtain accurate interaction energies, especially in regard to non-covalent interactions. However, they also have a detrimental impact on the sparsity of the one-particle density matrix (1-PDM), to a degree stronger than the spatial extent of the basis functions alone could explain. 
  This is despite the fact that the matrix elements of the 1-PDM of insulators (systems with significant HOMO-LUMO gaps)
  are expected to decay exponentially with increasing real-space distance from the diagonal
  and the asymptotic decay rate is expected to have a well-defined basis set limit.
  The observed low sparsity of the 1-PDM appears to be independent of representation and even persists after projecting the 1-PDM onto a real-space grid, 
  leading to the conclusion that this \enquote{curse of sparsity} is solely a basis set artifact, which, counterintuitively, becomes worse for larger basis sets, seemingly contradicting the notion of a well-defined basis set limit.
  We show that this is a consequence of the low locality of the contra-variant basis functions as quantified by the inverse overlap matrix $\mathbf{S}^{-1}$ being significantly less sparse than its co-variant dual.
  Introducing the model system of an infinite non-interacting chain of helium atoms, 
  we are able to quantify the exponential decay rate to be proportional to the diffuseness 
  as well as local incompleteness of the basis set, meaning small and diffuse basis sets are affected the most.
  Finally, we propose one solution to the conundrum in the form of the complementary auxiliary basis set (CABS) singles correction 
  in combination with compact, low l-quantum-number basis sets, showing promising results for non-covalent interactions.
\end{abstract}

\maketitle 
\section{Introduction}
The prospects of linear-scaling electronic structure theory have fascinated the electronic structure theory community
ever since its popularization by Kohn's work on the electronic \enquote{nearsightedness} principle\cite{kohn_density_1996}.
In particular, a sparse one-particle density matrix (1-PDM) with an asymptotically linear scaling number of significant elements --
either directly or indirectly through orbital localization --
lays the foundation of many low-scaling theories, such as 
linear scaling Fock/Kohn-sham-builds,\cite{burant_linear_1996,schwegler_linear_1997,ochsenfeld_linear_1998,neese_efficient_2009,laqua_efficient_2018,laqua_highly_2020,helmich-paris_improved_2021}
linear scaling self-consistent field (SCF) diagonalization alternatives\cite{challacombe_simplified_1999,niklasson_nonorthogonal_2005,jordan_comparison_2005,rubensson_density_2008,suryanarayana_optimized_2013},
as well as low-scaling variants of the 
M\o ller-Plesset perturbation theory (MP2),\cite{lambrecht_rigorous_2005,maurer_efficient_2013,glasbrenner_efficient_2020,wang_sparsity_2023,shi_local_2024}, 
Random-Phase-Approximation (RPA)\cite{schurkus_communication_2016,luenser_vanishing-overhead_2017,graf_accurate_2018,drontschenko_lagrangian-based_2021} 
and coupled cluster (CC) theory.\cite{meyer_pno-ci_1973,ahlrichs_pno_1975,pulay_localizability_1983,schutz_low-order_1999,schuetz_new_2002,neese_efficient_2009-1,riplinger_efficient_2013,schwilk_scalable_2017,sacchetta_effective_2022}

However, these low-scaling approaches struggle when employing large and especially diffuse basis sets.
This manifests as a late onset of the low-scaling regime,
larger (and sometimes erratic) cutoff errors from sparse treatment, or both.\cite{laqua_highly_2020,sandler_accuracy_2021,gray_assessing_2024}
Therefore, we decided to investigate the effect of basis sets onto the locality of the 1-PDM
and stumbled over the \enquote{conundrum} of diffuse basis sets:
On one side, they are necessary for accuracy (the blessing of accuracy),
yet on the other side, they are devastating for sparsity (the curse of sparsity).

We introduce this conundrum in sec.~\ref{sec:diffuse},
where we explore both curse and blessing for systems of practical relevance.
We then dive deeper into the root cause of the \enquote{curse of sparsity} by investigating the real-space 1-PDM,
thereby removing the non-uniqueness of basis set representations,
in sec.~\ref{sec:rs-1rdm}, followed
by a comparison of the locality between co- and contra-variant representations in sec.~\ref{sec:co_contra}.
We finalize this investigation with a mathematical analysis of a simple model system 
-- an idealized, non-interacting, infinite chain of helium atoms -- 
in sec.~\ref{sec:he}.
This model illuminates the mechanisms that may lead to non-locality 
by illustrating that non-locality may even arise in systems with highly local electronic structures
and basis sets that only consider nearest-neighbor overlap.
Putting all those insights together leads to a detailed characterization of the \enquote{curse of sparsity}
which is provided in sec.~\ref{sec:understanding}.
Finally, we propose a possible solution in
the form of the complementary auxiliary basis set (CABS) singles correction together with compact, l-quantum-number reduced basis sets
in sec.~\ref{sec:cabs}.

\section{The conundrum of diffuse basis sets} 
\label{sec:diffuse}
\subsection{The curse of sparsity}\label{sec:curse}
\begin{figure}[htbp]
  \begin{subfigure}[t]{0.24\textwidth}
    \includegraphics[width=0.95\textwidth]{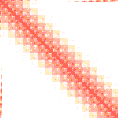}
    \caption{$\mathbf{P}$ (STO-3G)}
  \end{subfigure}
  \begin{subfigure}[t]{0.24\textwidth}
    \includegraphics[width=0.95\textwidth]{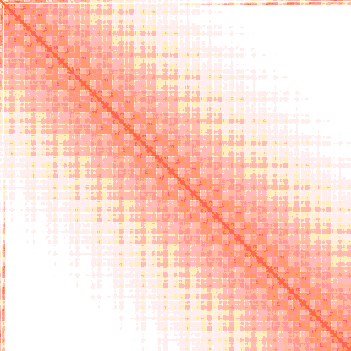}
    \caption{$\mathbf{P}$ (def2-SVP)}
  \end{subfigure}
  \begin{subfigure}[t]{0.24\textwidth}
    \includegraphics[width=0.95\textwidth]{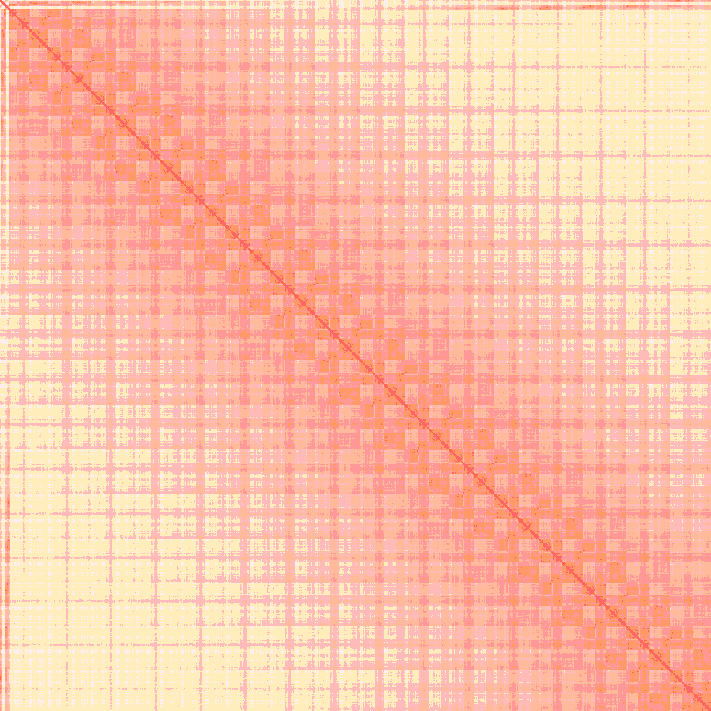}
    \caption{$\mathbf{P}$ (def2-TZVP)
  \label{fig:dna_P_tzvp}
    }
  \end{subfigure}
  \begin{subfigure}[t]{0.24\textwidth}
    \includegraphics[width=0.95\textwidth]{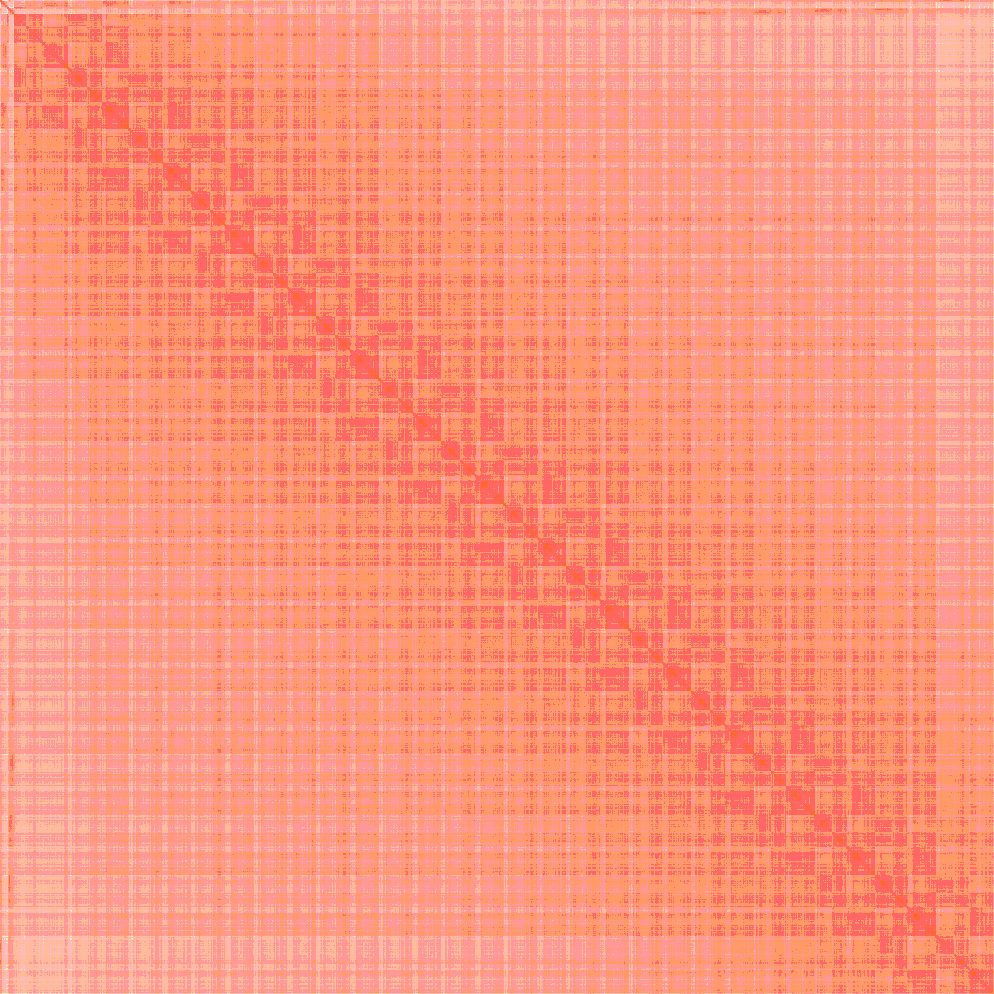}
    \caption{$\mathbf{P}$ (def2-TZVPPD)}
  \end{subfigure}
  \caption{
    Block-sparsity-pattern (32$\times$32 blocks) of the converged Hartree-Fock SCF one-particle density matrix for \ch{(AT)16} DNA-strand (1052 atoms).
    White pixels correspond to blocks with (per element averaged) $L_2$ norms of $<$\num{e-10}. 
  \label{fig:dna_P}
}
\end{figure}

The detrimental impact of diffuse basis functions (sometimes referred to as augmentation functions) onto the sparsity of the SCF-converged 1-PDM is illustrated in \cref{fig:dna_P} at
the example of a DNA-fragment comprising 16 base pairs totaling 1052~atoms.
This should represent a prototypical example for Kohn's \enquote{nearsightedness} principle.
Unfortunately, while there is significant sparsity for small basis sets (especially STO-3G), 
even the medium sized diffuse basis set def2-TZVPPD removes essentially all usable sparsity, i.e., nearly all off-diagonal elements of the 1-PDM are too significant to be discarded.
This detrimental impact on electronic sparsity is what we refer to as the \enquote{curse} of diffuse basis functions
and represents the main focus of this work.

\subsection{The blessing of accuracy} \label{sec:blessing}
Although larger basis sets in general result in somewhat deteriorated sparsity as apparent from the def2-TZVP results in \cref{fig:dna_P_tzvp}),
the problem is mostly associated with diffuse basis sets.
Therefore, the obvious solution seems to be to simply avoid these basis sets.
Unfortunately, it was shown in many studies
that augmentation with diffuse functions is absolutely essential for an accurate description of non-covalent interactions (NCIs).\cite{marshall_basis_2011,papajak_perspectives_2011,witte_push_2016,nagy_approaching_2019,gray_comprehensive_2022} 

We demonstrate this effect here with computations on the ASCDB benchmark,\cite{morgante_statistically_2019} which contains a statistically relevant cross section of relative energies on a wide range of chemical problems. As an example method, we chose the range-separated hybrid density functional $\omega$B97X-V\cite{mardirossian_b97x-v_2014} and illustrated its accuracy in combination with both augmented as well as unaugmented basis sets from the Karlsruhe family (def2-X)\cite{weigend_balanced_2005} as well as from Dunning's correlation-consistent basis sets (cc-pVXZ).\cite{dunning_gaussian_1989,woon_gaussian_1993,woon_gaussian_1994,balabanov_systematically_2005} 
All basis sets were used as provided by the Basis Set Exchange.\cite{pritchard_new_2019} The resulting root mean-square deviations (RMSD) for the entire benchmark as well as just the NCIs are given in \cref{tab:basisset_errors}.

\begin{table}[htbp]
  \caption{Combined Method ($\omega$B97X-V)
  and Basis set errors (M+B) and only basis set error (B) [\si{\kilo\cal\per\mol}] referenced to aug-cc-pV6Z as root-mean-square-deviation (RMSD) 
  for full ASCDB benchmark\cite{morgante_statistically_2019} (RMSD) and only non-covalent interactions (NCI RMSD).
  No counterpoise corrections are applied. 
  Timings are given in seconds for one full $\omega$B97X-V SCF calculation for an \ch{(AT)4}-DNA fragment (260~atoms).
  Additional computational details are provided in appendix \ref{sec:comp_details}.
  \label{tab:basisset_errors}
  }
  \begin{ruledtabular}
  \begin{tabular}{lrrrrrrrr}
    Basis set & \multicolumn{1}{c}{RMSD (B)} & \multicolumn{1}{c}{RMSD (M+B)} & \multicolumn{1}{c}{NCI RMSD (B)} & \multicolumn{1}{c}{NCI RMSD (M+B)} & Time [s]\\
    \colrule
    def2-SVP    & 30.84 & 33.32 & 31.33 & 31.51 &   151\\
    def2-TZVP   &  5.50 & 17.36 &  7.75 &  8.20 &   481\\
    def2-QZVP   &  1.93 & 16.53 &  1.73 &  2.98 &  1935\\
    cc-pVDZ     & 25.34 & 32.82 & 30.17 & 30.31 &   178\\
    cc-pVTZ     &  9.13 & 18.52 & 12.46 & 12.73 &   573\\
    cc-pVQZ     &  4.37 & 16.99 &  5.69 &  6.22 &  1773\\
    cc-pV5Z     &  1.28 & 16.46 &  1.40 &  2.81 &  6439\\
    cc-pV6Z     &  0.47 & 16.49 &  0.41 &  2.47 & 15265\\
    def2-SVPD   & 23.45 & 26.50 &  7.04 &  7.53 &   521\\
    def2-TZVPPD &  1.82 & 16.40 &  0.73 &  2.45 &  1440\\
    def2-QZVPPD &  0.62 & 16.69 &  0.33 &  2.40 &  3415\\
    aug-cc-pVDZ & 15.94 & 26.75 &  4.32 &  4.83 &   975\\
    aug-cc-pVTZ &  3.90 & 17.01 &  1.23 &  2.50 &  2706\\
    aug-cc-pVQZ &  1.78 & 16.90 &  0.61 &  2.40 &  7302\\
    aug-cc-pV5Z &  0.32 & 16.57 &  0.09 &  2.39 & 24489\\
    aug-cc-pV6Z &  -    & 16.57 & -     &  2.41 & 57954\\
  \end{tabular}
  \end{ruledtabular}
\end{table}

The fact that def2-TZVPPD and aug-cc-pVTZ are the smallest basis sets where the combined method and basis error for NCIs (\SI{2.45}{\kilo\cal\per\mol} and \SI{2.50}{\kilo\cal\per\mol} respectively) are sufficiently converged compared to the aug-cc-pV6Z results of \SI{2.41}{\kilo\cal\per\mol}, 
confirms the necessity for diffuse basis functions.
Instead, without augmentation, only cc-pV6Z yields satisfactory accuracy and no unaugmented basis set from the Karlsruhe (def2-) family achieves similar quality. 
While the addition of diffuse basis functions adds substantial computational costs,
especially for larger system like the here employed example of a 260~atoms \ch{(AT)4} DNA-fragment
(e.g. \SI{1440}{\second} for def2-TZVPPD vs.~\SI{481}{\second} for def2-TZVP),
the gain in accuracy outweighs the additional cost, e.g.~ def2-TZVPPD is over 10 times faster as cc-pv6z 
(\SI{1440}{\second} vs.~\SI{15265}{\second})
while resulting in similar overall accuracy (combined method and basis error).
Clearly, diffuse basis sets are essential to obtain accurate NCI energies at reasonable computational costs. 
This is what we refer to as the \enquote{blessing} of diffuse basis functions.

The core of the conundrum emerges when considering large, non-trivial chemical environments. Important examples thereof are solvent/protein cavities, 
solid state surfaces, or supramolecular host-guest complexes, 
where electronic sparsity is highly desirable in order to manage the computational complexity of such large systems but
where NCIs also play a decisive role in the description of properties and reactivity. A conventional way of tackling this issue is the use of counter-poise correction, e.g.~with the Boys-Bernardi method.\cite{boys_calculation_1970}
Nonetheless, Boys-Bernardi counter-poise correction,
which was indeed shown to provide significant improvements for NCIs,\cite{van_duijneveldt_state_1994,marshall_basis_2011,witte_push_2016,kesharwani_s66_2018,gray_systematic_2022}
is rarely applicable to these situations in practice,
because the necessary fragmentation scheme
requires \emph{inter}-molecular NCIs 
while being incompatible with \emph{intra}-molecular interactions.

We thus decided to shed more light on the \enquote{cursed} side of the conundrum and try to answer the following question:
Why are diffuse basis functions so detrimental to electronic sparsity -- 
much more than the increased extent and basis-function overlap could explain?
This is particularly puzzling, since the apparent non-locality is in stark contrast with the well-studied 
asymptotically exponential decay of the 1-PDM that is at the foundation of low-scaling electronic structure methods.\cite{kohn_density_1996, baer_sparsity_1997,maslen_locality_1998,ismail-beigi_locality_1999,jedrzejewski_exact_2004,rubensson_bringing_2011}

\section{The One-Particle Density Matrix in Real-Space Representation}\label{sec:rs-1rdm}
In order to solve this riddle of non-locality, we decided to remove the basis-set representation itself as much as possible from our analysis and
focus on the real-space 1-PDM $\rho(\mathbf{r},\mathbf{r}')$ instead of the atomic orbital (AO) 1-PDM $P_{\mu\nu}$.
The real-space 1-PDM is easily computed
for a grid of coordinates $\mathbf{r}$ and $\mathbf{r}'$ as
\begin{equation}
  \rho(\mathbf{r},\mathbf{r}') = \sum_{\mu\nu} P_{\mu\nu} \chi_\mu(\mathbf{r})\chi_\nu(\mathbf{r}')
\end{equation}
where $\chi_\mu(\mathbf{r})$ denote co-variant AO basis functions.

\begin{figure}[htbp]
  \begin{subfigure}[t]{0.24\textwidth}
    \includegraphics[width=0.95\textwidth]{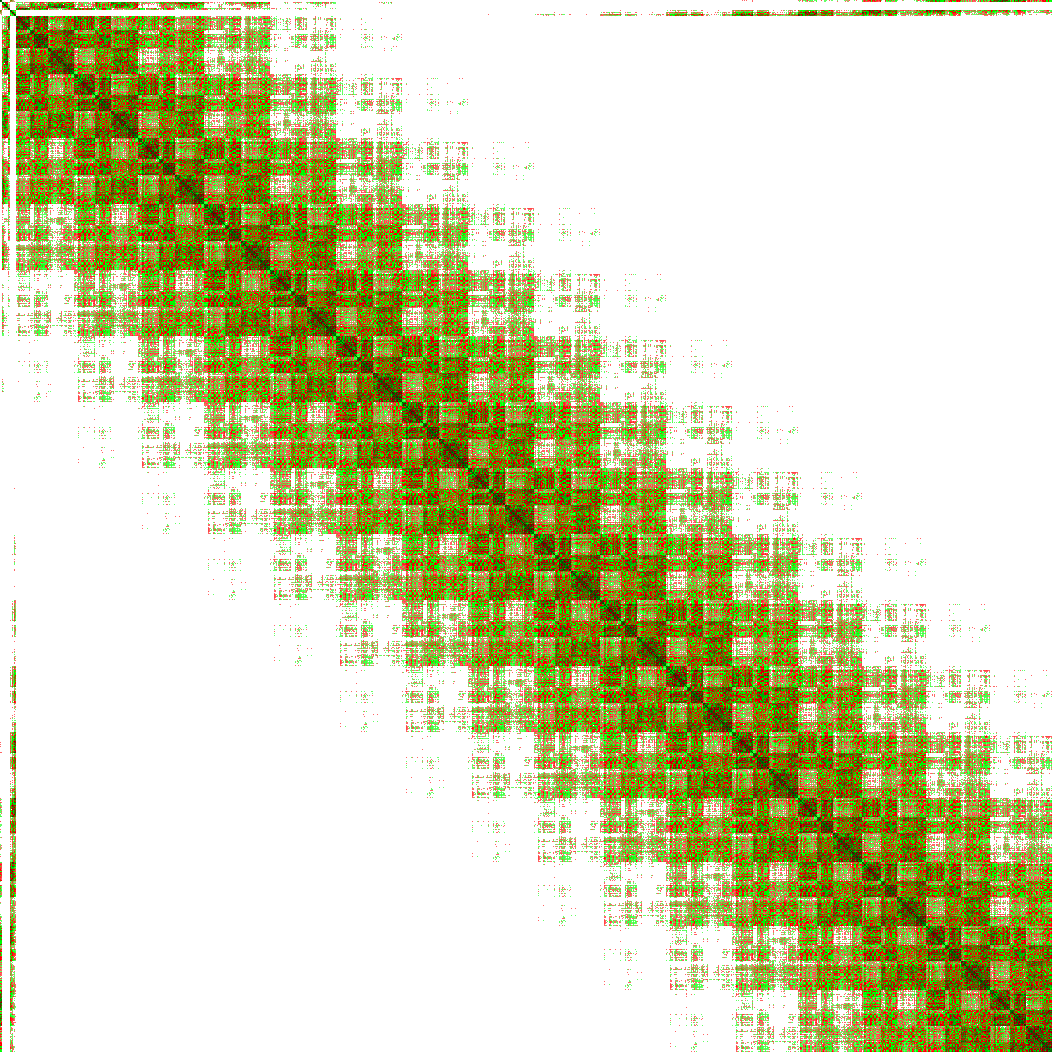}
    \caption{$\rho(\mathbf{r},\mathbf{r}')$ (STO-3G)}
  \end{subfigure}
  \begin{subfigure}[t]{0.24\textwidth}
    \includegraphics[width=0.95\textwidth]{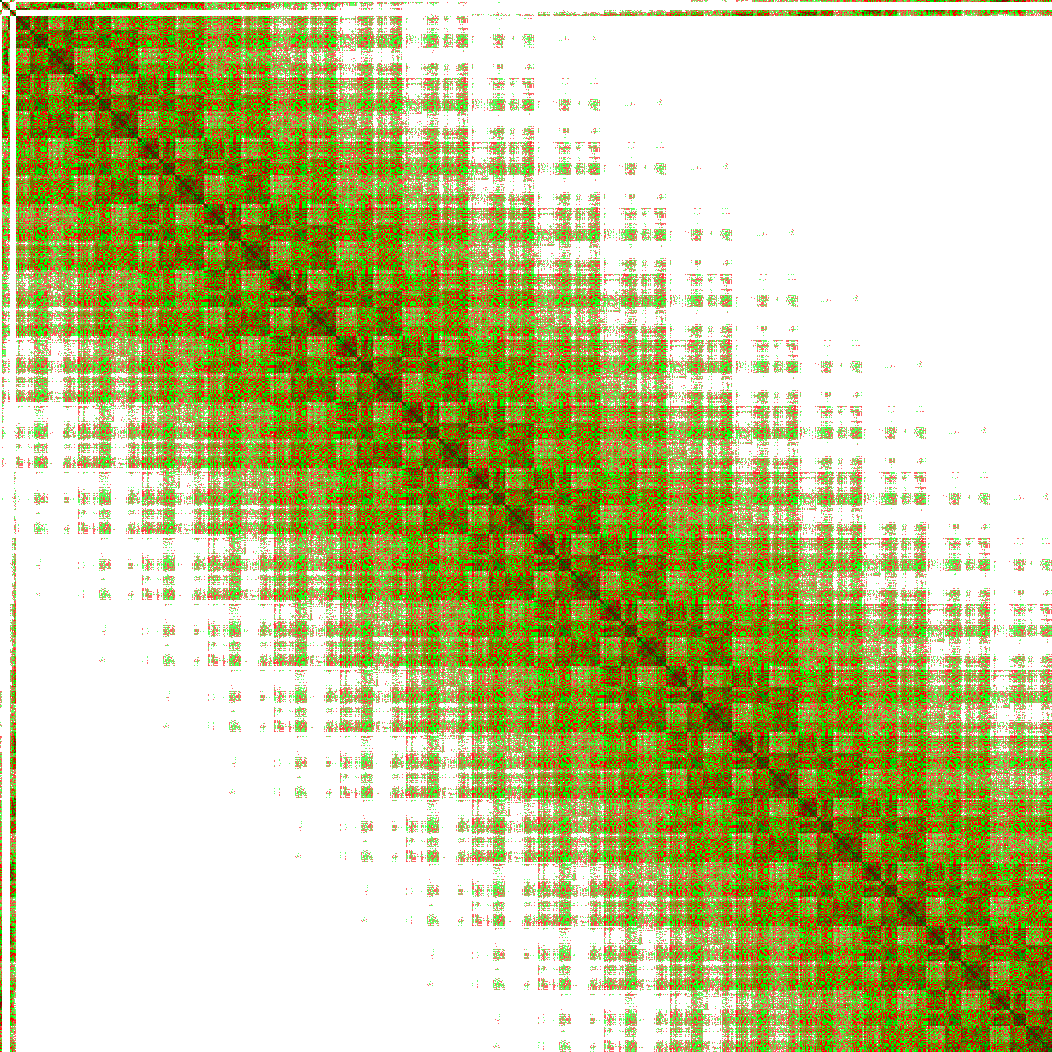}
    \caption{$\rho(\mathbf{r},\mathbf{r}')$ (def2-SVP)}
  \end{subfigure}
  \begin{subfigure}[t]{0.24\textwidth}
    \includegraphics[width=0.95\textwidth]{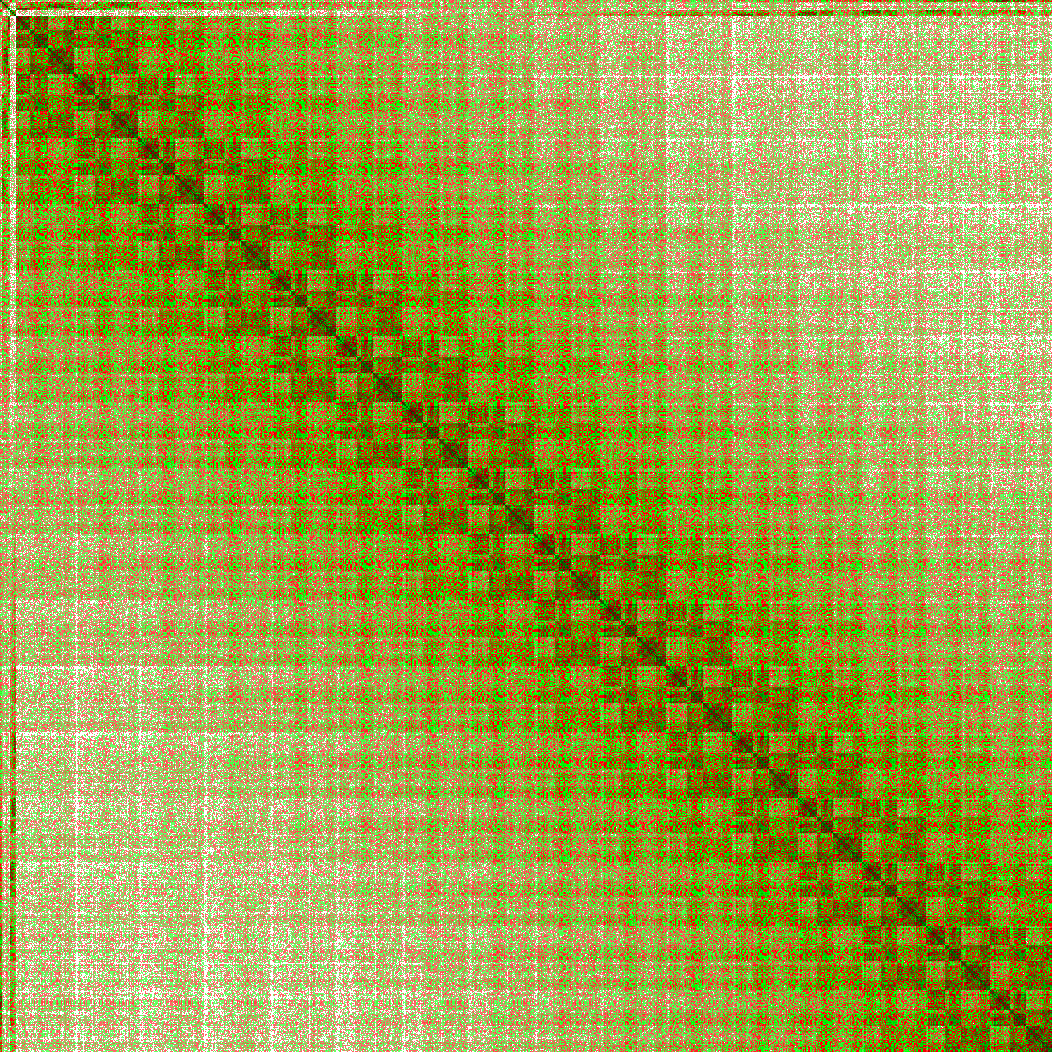}
    \caption{$\rho(\mathbf{r},\mathbf{r}')$ (def2-TZVP)}
  \end{subfigure}
  \begin{subfigure}[t]{0.24\textwidth}
    \includegraphics[width=0.95\textwidth]{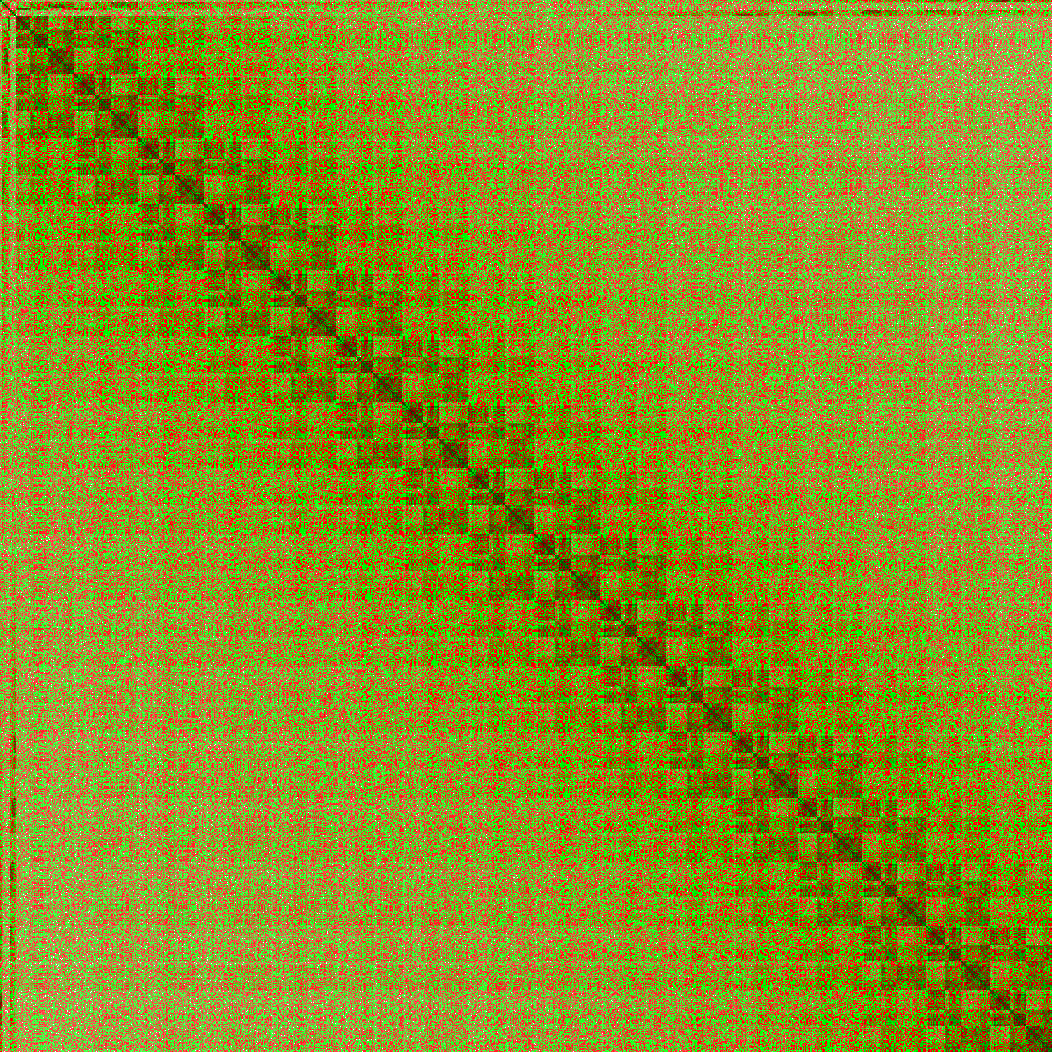}
    \caption{$\rho(\mathbf{r},\mathbf{r}')$ (def2-TZVPPD)}
  \end{subfigure}
  \caption{
    Element-wise sparsity-pattern 
    (each pixel corresponds to one matrix element; shades of green: positive values, shades of red: negative values, white: absolute values $<$\num{e-10})
    of real-space density matrix $\rho(\mathbf{r},\mathbf{r}')$ for \ch{(AT)16} DNA-strand (1052 atoms).
    One grid-point per atom, each at \SI{1}{a.u.} distance from parent nucleus.
  \label{fig:dna_P_rs}
}
\end{figure}

The results, depicted in \cref{fig:dna_P_rs}, are truly intriguing:
The 1-PDM obtained from a SCF calculation employing a diffuse basis sets (def2-TZVPPD as the extreme case) 
is non-local even in the most local of representations -- real-space grids!

\begin{figure}[htbp]
    \includegraphics[width=0.48\textwidth]{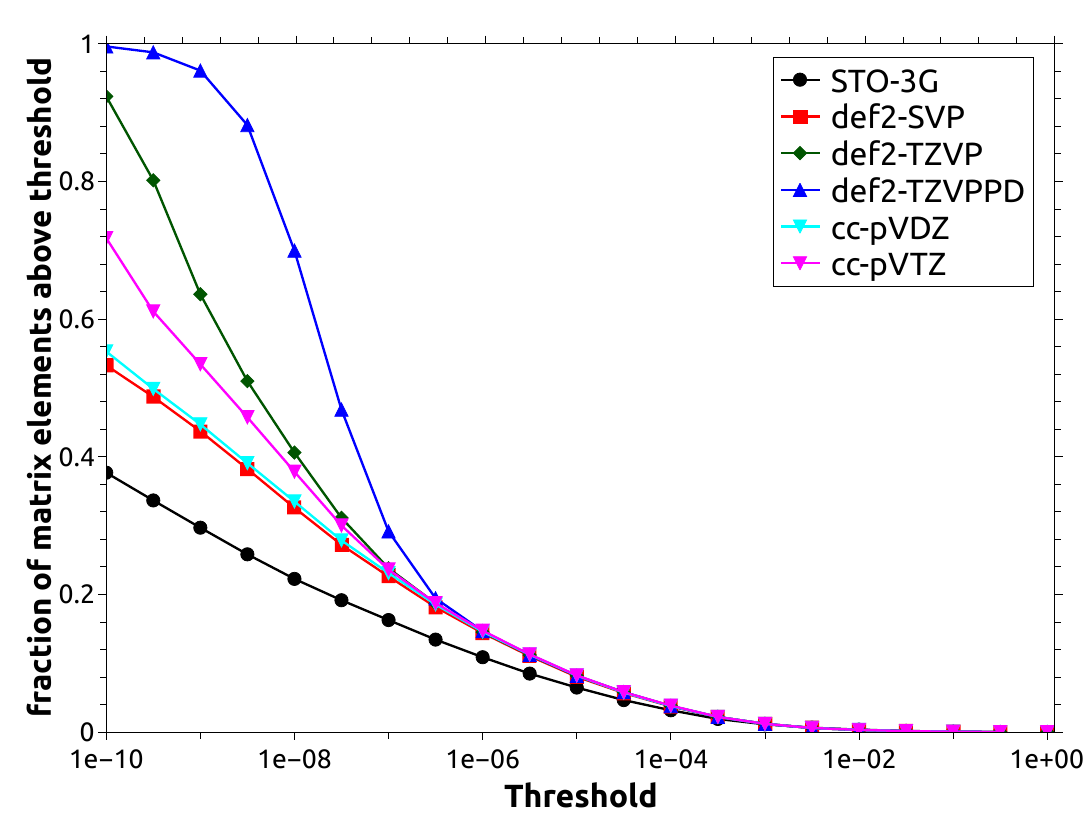}
  \caption{
    Element-wise sparsity counts of real-space density matrix $\rho(\mathbf{r},\mathbf{r}')$ for \ch{(AT)16} DNA-strand (1052 atoms).
    One grid-point per atom, each at \SI{1}{a.u.} distance from parent nucleus.
  \label{fig:dna_P_rs_count}
}
\end{figure}

In order to shed more light on the existence or absence of sparsity, we count the number of significant elements as shown in \cref{fig:dna_P_rs_count}. 
Interestingly, this reveals that all 
basis sets except for STO-3G result in nearly identical locality up to a threshold of about \num{e-6}
and only start to differ significantly between basis sets for tighter thresholds.
Since we expect values of this magnitude to only occur in the far tail of the 1-PDM, this result indicates that there might be an ill-defined basis set limit within that regime. 
Evidently, an increase in basis set size and especially augmentation with diffuse functions 
does not converge to well-defined basis set limits for the asymptotic decay rate of $\rho(\mathbf{r},\mathbf{r}')$
and instead just results in slower and slower decay for larger and more diffuse the basis sets.
This is troublesome given that $\rho(\mathbf{r},\mathbf{r}')$ -- unlike its AO analogue $P_{\mu\nu}$ --
should be independent of representation and thus possess a well-defined complete basis set limit.

\begin{figure}[htbp]
  \begin{subfigure}[t]{0.48\textwidth}
    \includegraphics[width=0.99\textwidth]{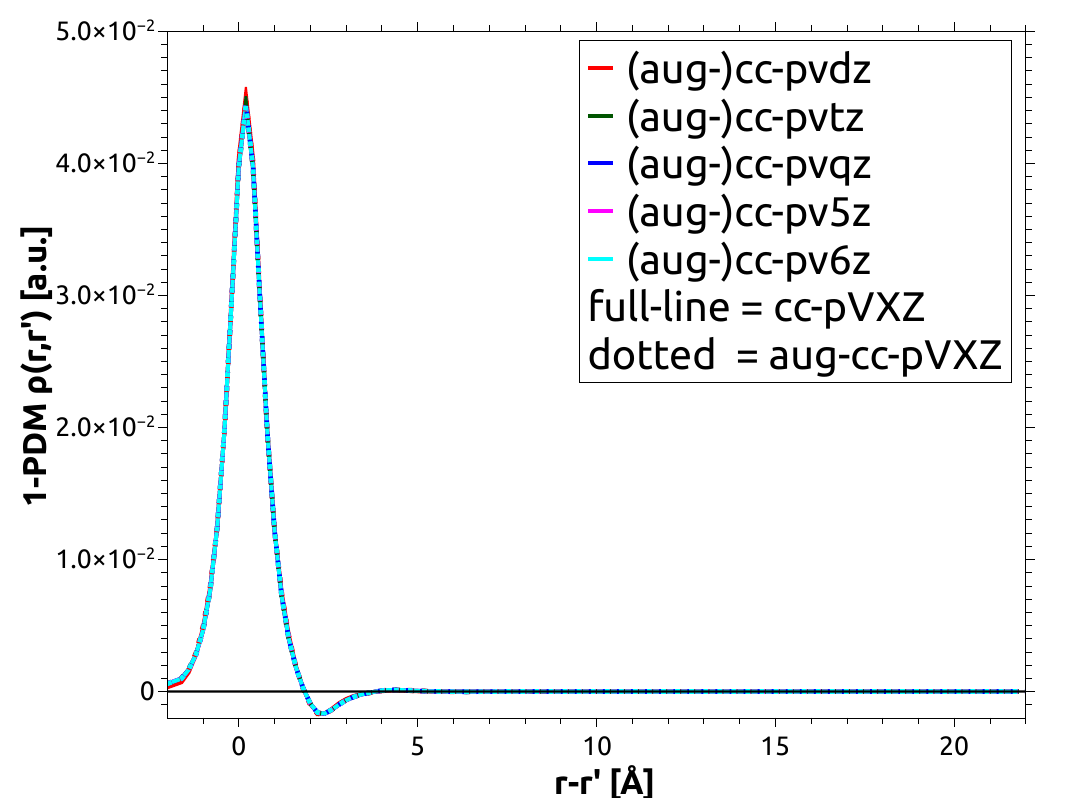}
    \caption{\label{fig:He10_dna_P_rs_a}}
  \end{subfigure}
  \begin{subfigure}[t]{0.48\textwidth}
    \includegraphics[width=0.99\textwidth]{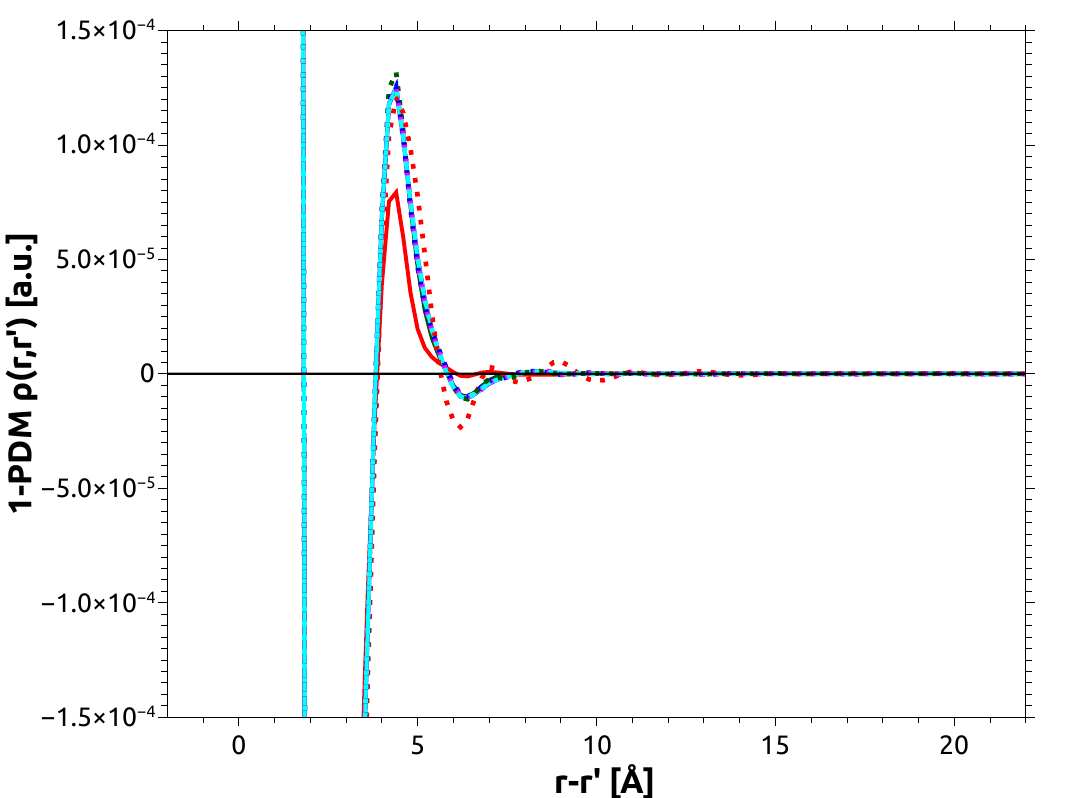}
    \caption{\label{fig:He10_dna_P_rs_b}}
  \end{subfigure}
  \begin{subfigure}[t]{0.48\textwidth}
    \includegraphics[width=0.99\textwidth]{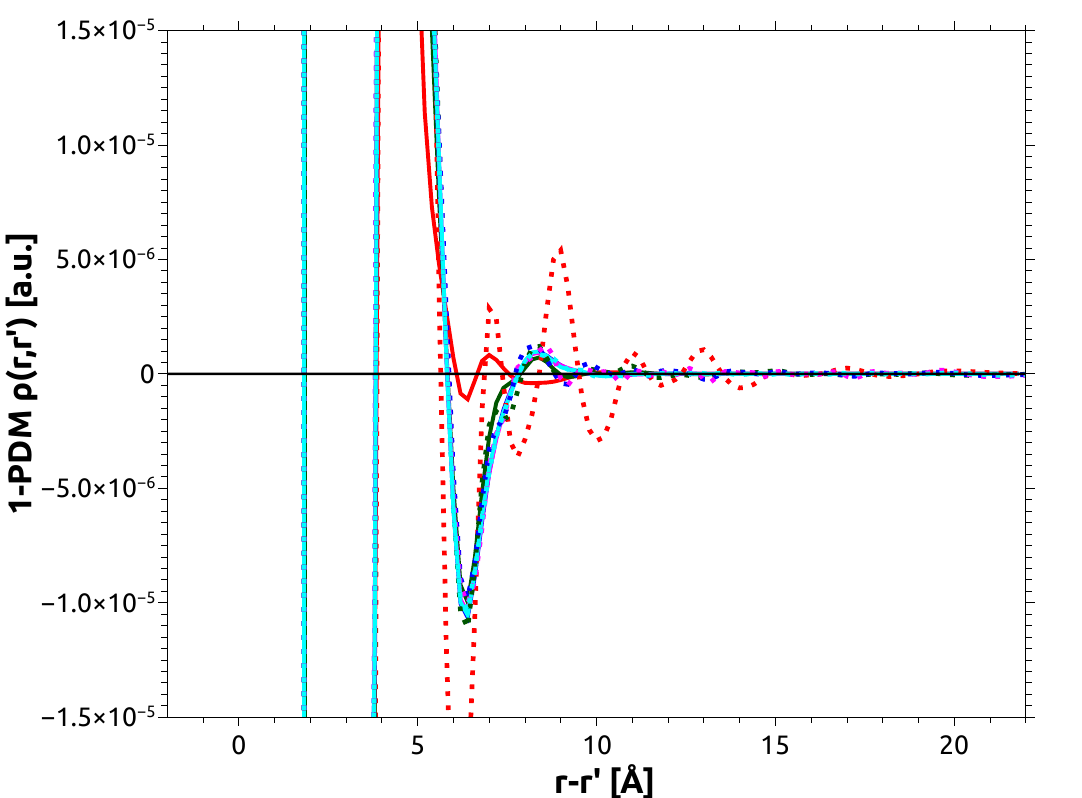}
    \caption{\label{fig:He10_dna_P_rs_c}}
  \end{subfigure}
  \begin{subfigure}[t]{0.48\textwidth}
    \includegraphics[width=0.99\textwidth]{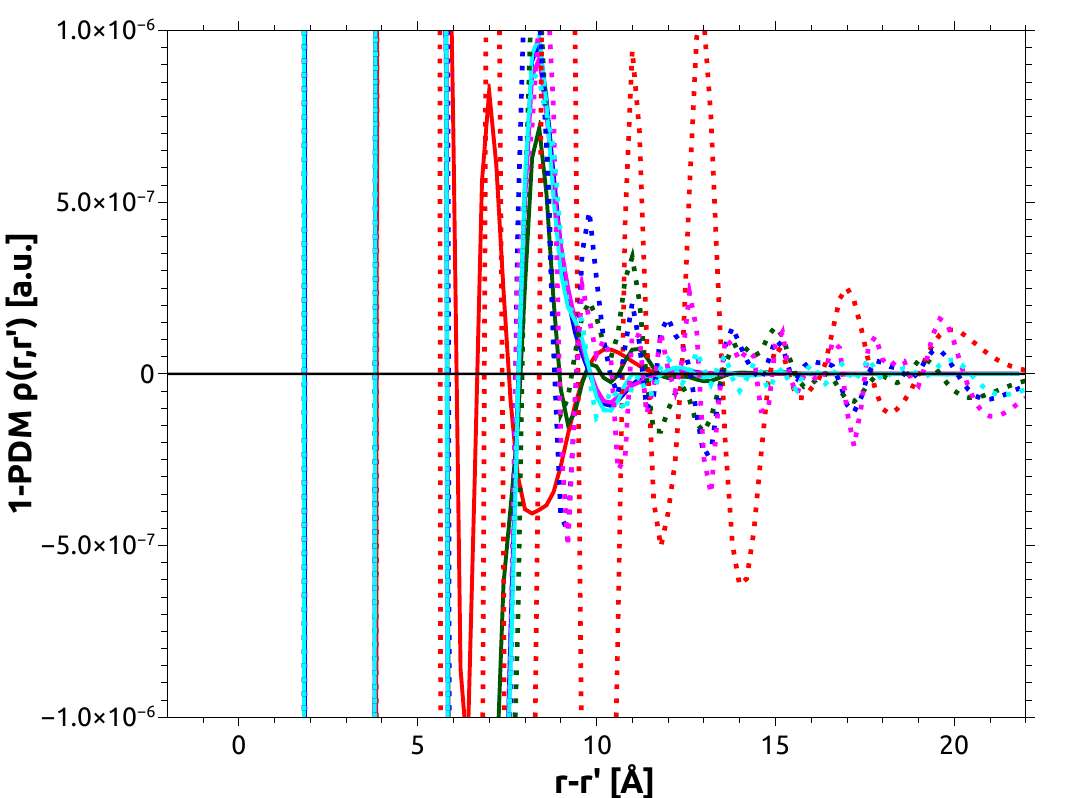}
    \caption{\label{fig:He10_dna_P_rs_d}}
  \end{subfigure}
  \caption{
    One row ($\mathbf{r}$ fixed) of real-space 1-PDM $\rho(\mathbf{r},\mathbf{r}')$ 
    for linear \ch{He10}-chain (\ch{He}-\ch{He} distance of \SI{2.0}{\angstrom}). 
    Real-space grid points parallel to chain of nuclei at a constant distance of \SI{1}{a.u.}.
    Figures~a-d differ only by level of magnification.
  \label{fig:He10_dna_P_rs}
}
\end{figure}

We furthermore explore the ill-definition of the complete basis set limit in \cref{fig:He10_dna_P_rs}
where the off-diagonal elements of the real-space 1-PDM $\rho(\mathbf{r},\mathbf{r}')$ 
are depicted for a linear helium chain -- 
a prototypical example of a strong insulator where a very sparse 1-PDM is expected.
Indeed, the initial decay to about \num{e-4} is rapid and identical for all tested basis sets 
just like in the \ch{(AT)16} DNA-strand. 
However, in the magnified plots in \cref{fig:He10_dna_P_rs_c,fig:He10_dna_P_rs_d} the
influence of diffuse functions becomes very clear:
All of the augmented basis sets (aug-cc-pVXZ) exhibit artificial oscillations on the order of \numrange{e-7}{e-6},
whereas none of the non-augmented basis (cc-pVXZ) show such a behavior and instead continue to decay exponentially, just as expected.
Surprisingly, the smallest augmented basis set aug-cc-pVDZ leads to the largest oscillations by far.

The great agreement between all non-augmented (cc-pVXZ) basis sets 
and the fact that there seems to be a systematic convergence when increasing the size from cc-pVDZ to cc-pV6Z
indicates that in the asymptote of the off-diagonal tail of $\rho(\mathbf{r},\mathbf{r}')$,
cc-pV6Z and not its augmented counterpart aug-cc-pV6Z provides the best approximation to 
the true basis set limit.
This is all the more astonishing given that the aug-cc-pVXZ basis sets are true supersets of their respective cc-pVXZ counterpart.

\section{Locality of co- vs. contravariant basis functions}\label{sec:co_contra}
Overall, these artificial oscillations seem to be the root cause of the non-sparsity \enquote{curse} 
of diffuse (and generally larger) basis sets. 
To our best knowledge, a reason for the divergence of the far off-diagonal of the real-space 1-PDM from the true basis set limit if augmented basis functions are employed has not been given or hypothesised in the literature.
The effect of absent sparsity is reminiscent of a typical overfitting error, 
where too many degrees of freedom lead to artifacts in the optimal output due to underdetermination. 
Here, the optimization function is given by the SCF energy functional, but what exactly is the parameter set which the 1-PDM is represented in?

\begin{figure}[htbp]
  \begin{subfigure}[t]{0.24\textwidth}
    \includegraphics[width=0.95\textwidth]{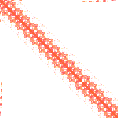}
    \caption{$\mathbf{S}$ (STO-3G)
    \label{fig:dna_S_sto3g}
    }
  \end{subfigure}
  \begin{subfigure}[t]{0.24\textwidth}
    \includegraphics[width=0.95\textwidth]{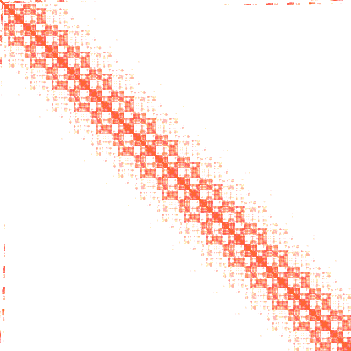}
    \caption{$\mathbf{S}$ (def2-SVP)
    \label{fig:dna_S_svp}
    }
  \end{subfigure}
  \begin{subfigure}[t]{0.24\textwidth}
    \includegraphics[width=0.95\textwidth]{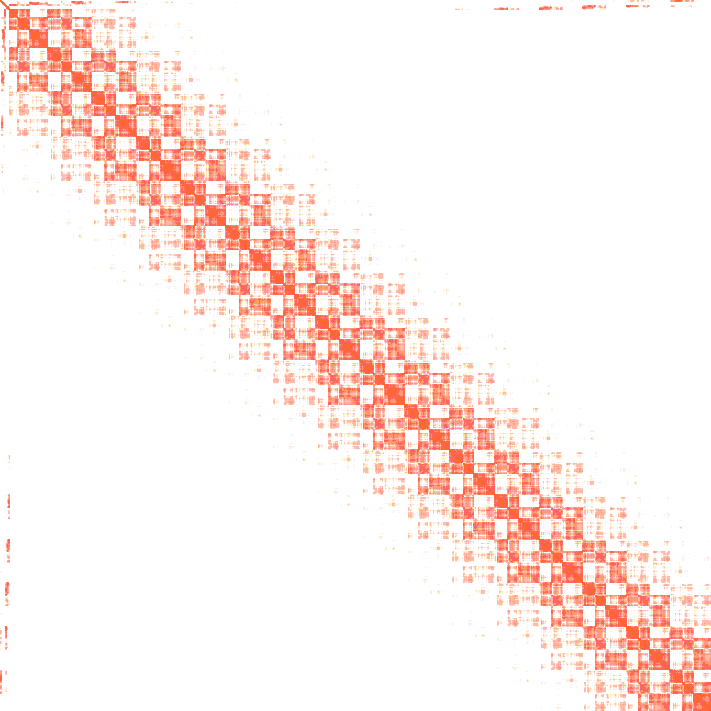}
    \caption{$\mathbf{S}$ (def2-TZVP)
    \label{fig:dna_S_tzvp}
    }
  \end{subfigure}
  \begin{subfigure}[t]{0.24\textwidth}
    \includegraphics[width=0.95\textwidth]{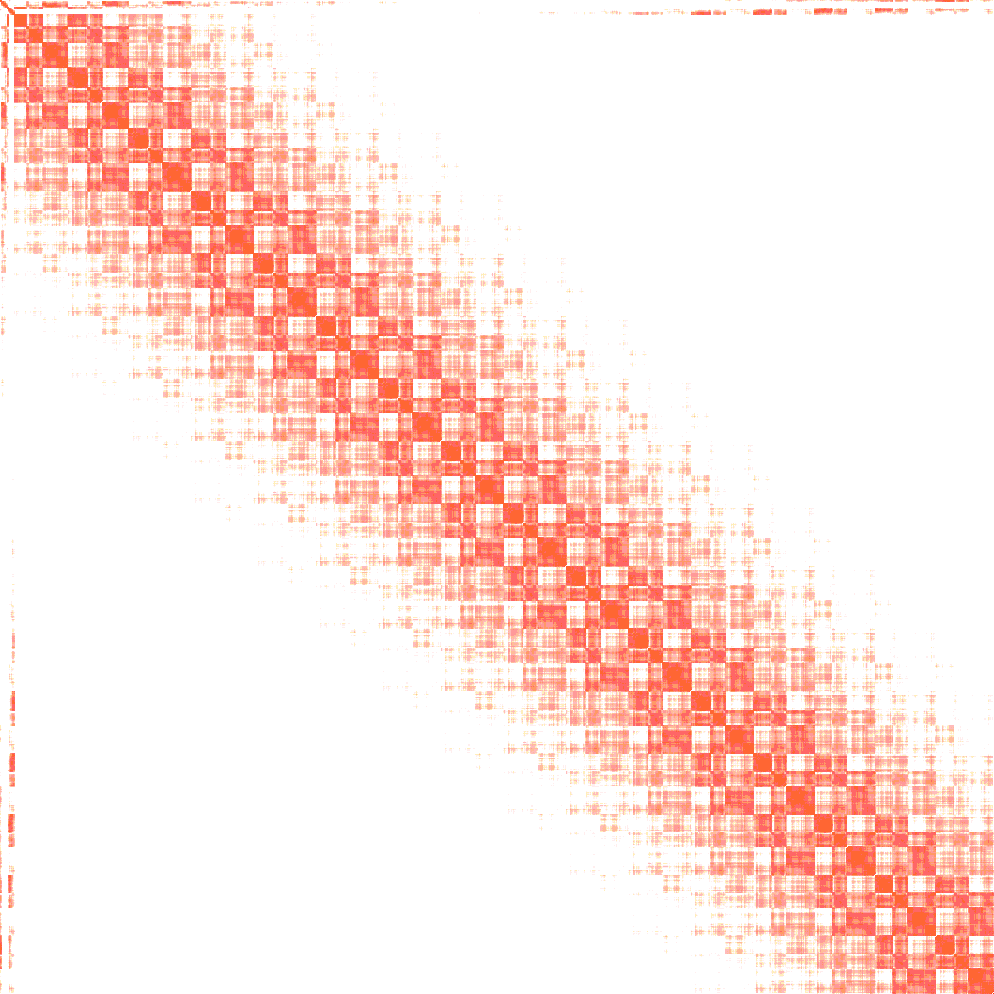}
    \caption{$\mathbf{S}$ (def2-TZVPPD)
    \label{fig:dna_S_tzvppd}
    }
  \end{subfigure}
  \begin{subfigure}[t]{0.24\textwidth}
    \includegraphics[width=0.95\textwidth]{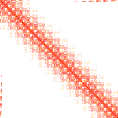}
    \caption{$\mathbf{S}^{-1}$ (STO-3G)
    \label{fig:dna_Sinv_sto3g}
    }
  \end{subfigure}
  \begin{subfigure}[t]{0.24\textwidth}
    \includegraphics[width=0.95\textwidth]{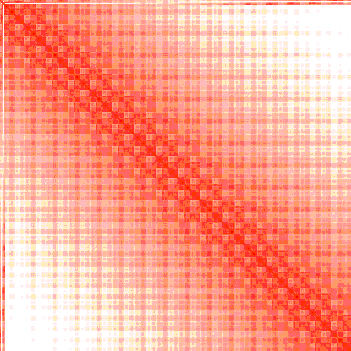}
    \caption{$\mathbf{S}^{-1}$ (def2-SVP)
    \label{fig:dna_Sinv_svp}
    }
  \end{subfigure}
  \begin{subfigure}[t]{0.24\textwidth}
    \includegraphics[width=0.95\textwidth]{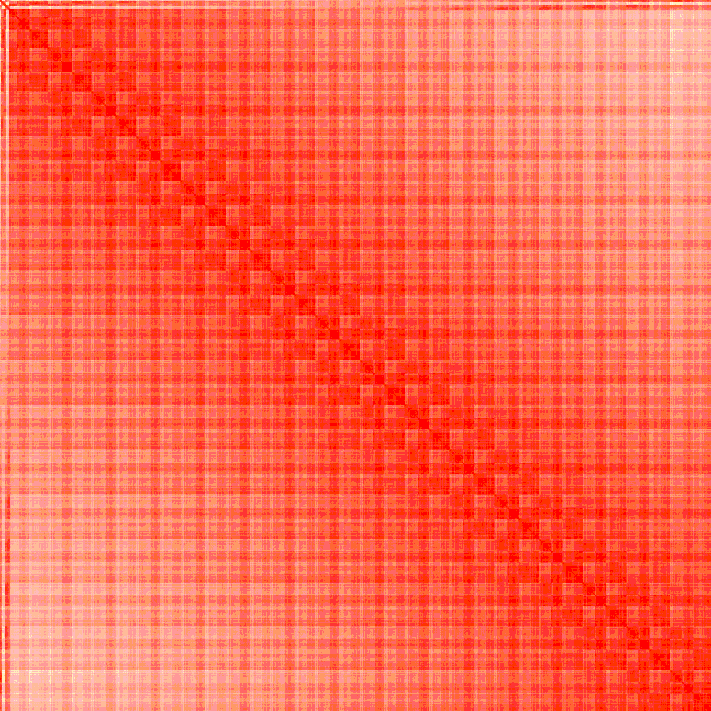}
    \caption{$\mathbf{S}^{-1}$ (def2-TZVP)
    \label{fig:dna_Sinv_tzvp}
    }
  \end{subfigure}
  \begin{subfigure}[t]{0.24\textwidth}
    \includegraphics[width=0.95\textwidth]{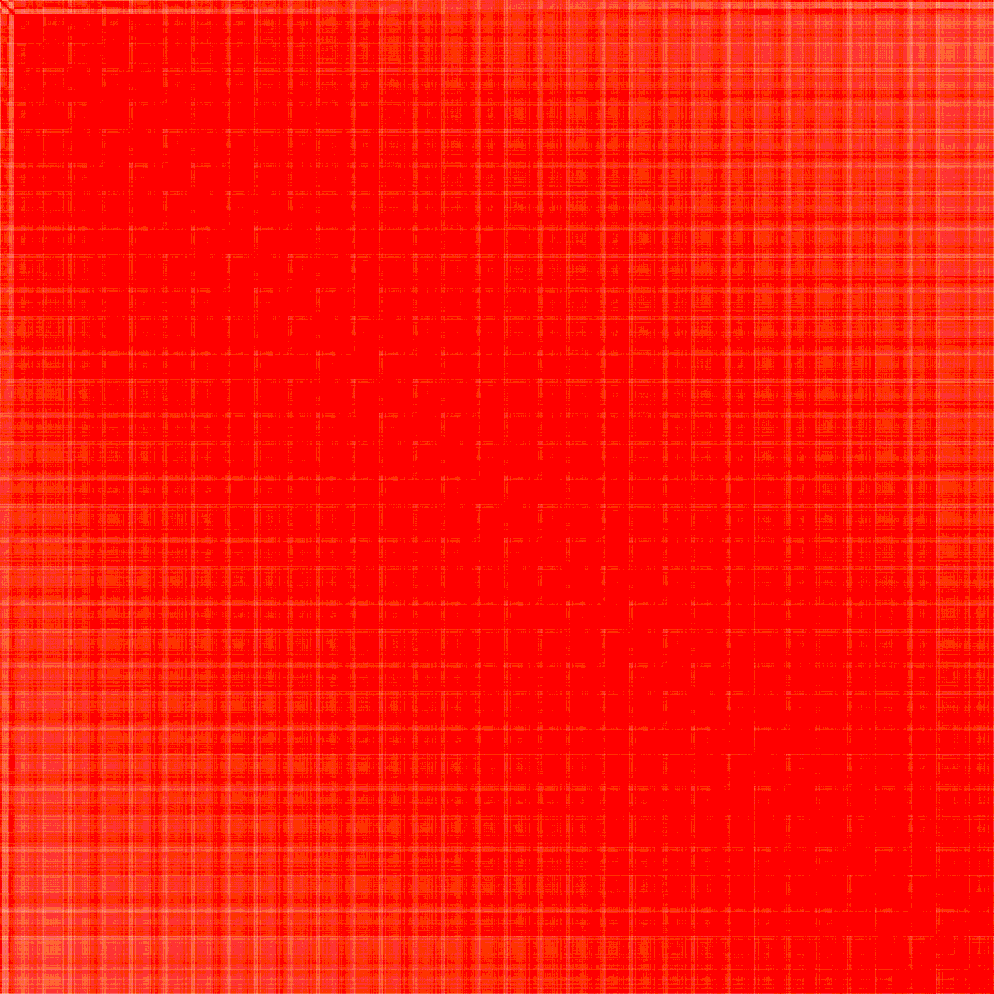}
    \caption{$\mathbf{S}^{-1}$ (def2-TZVPPD)
    \label{fig:dna_Sinv_tzvppd}
    }
  \end{subfigure}
  \caption{
    Block-sparsity-pattern (32$\times$32 blocks) of overlap matrix (a-d) and inverse overlap matrix (e-h)
    for \ch{(AT)16} DNA-strand (1052 atoms).
    White pixels correspond to blocks with (per element averaged) $L_2$ norms of $<$\num{e-10}. 
  \label{fig:dna_S_Sinv}
}
\end{figure}

Since the 1-PDM is the solution of a minimization problem, its AO representation 
\begin{equation}
  P_{\mu\nu} = \int \mathrm{d}\mathbf{r}\int \mathrm{d}\mathbf{r}'\tilde{\chi}_\mu(\mathbf{r})\rho(\mathbf{r},\mathbf{r}')\tilde{\chi}_\nu(\mathbf{r}')
\end{equation}
is the contra-variant\cite{head-gordon_tensors_2000} dual
\begin{equation}
  \tilde{\chi}_\mu (\mathbf{r}) = \sum_{\mu} (\mathbf{S}^{-1})_{\mu\nu} \chi_\nu(\mathbf{r})
\end{equation}
of the original AO basis set, requiring a transformation with the inverse overlap matrix $\mathbf{S}^{-1}$.
This suggests that, rather than the co-variant basis set 
(as measured by the overlap matrix $\mathbf{S}$ in \cref{fig:dna_S_sto3g,fig:dna_S_svp,fig:dna_S_tzvp,fig:dna_S_tzvppd}),
its contra-variant dual 
(as measured by the inverse overlap matrix $\mathbf{S}^{-1}$ in \cref{fig:dna_Sinv_sto3g,fig:dna_Sinv_svp,fig:dna_Sinv_tzvp,fig:dna_Sinv_tzvppd})
defines the parameter space for the optimization of the 1-PDM.

This hypothesis is trivially true in cases where the number of basis functions equals the number of electron pairs, i.e.~where there is no virtual space. An example of this is
an arbitrarily long helium chain with a minimal basis set such as STO-3G, where 
\begin{equation}\label{eq:P_min}
  \mathbf{P} = \mathbf{S}^{-1}
\end{equation}
follows directly from the well-known idempotency relation\cite{mcweeny_hartree-fock_1959}
\begin{equation}
  \mathbf{PSP} = \mathbf{P},
\end{equation}
since $\mathbf{P}$ is not rank-deficient in this special case.
We argue that in a more general case, $\mathbf{S}^{-1}$ can still be regarded as a diagnostic for the parameter space in which the 1-PDM is optimized in, 
meaning a very non-local $\mathbf{S}^{-1}$ 
(and consequently non-local contra-variant basis functions $\tilde{\chi}_\nu(\mathbf{r})$)
as observed in \cref{fig:dna_Sinv_tzvppd} 
may provide the parameter space that enables the observed overfitting in the off-diagonal 1-PDM.

Therefore, the observed similarity between the sparsity patterns of $\mathbf{S}^{-1}$ (\cref{fig:dna_Sinv_sto3g,fig:dna_Sinv_svp,fig:dna_Sinv_tzvp,fig:dna_Sinv_tzvppd}) and $\mathbf{P}$ (\cref{fig:dna_P}) is arguably not just coincidental, but causal --
a non-local $\mathbf{S}^{-1}$ can (and in practice often will) lead to a non-local $\mathbf{P}$, 
even if the corresponding co-variant overlap matrix $\mathbf{S}$ (\cref{fig:dna_S_sto3g,fig:dna_S_svp,fig:dna_S_tzvp,fig:dna_S_tzvppd})
is still sufficiently local.

\section{The infinite Helium chain}\label{sec:he}
So how precisely does the non-locality in $\mathbf{S}^{-1}$ come about and how does it affect the locality of $\mathbf{P}$?
To answer these questions and to quantify the dependence of the decay rate of the real-space 1-PDM on the choice of basis set, we construct a simple model system comprised of an idealized, infinite,
non-interacting chain of helium-like atoms. In this system, the exact real-space 1-PDM is atom-block diagonal,
i.e., we can uniquely associate real-space coordinates $\mathbf{r}$ to a parent atom $A$.
Moreover, real-space 1-PDM matrix elements $\rho(\mathbf{r},\mathbf{r}')$ are only non-zero if $\mathbf{r}$ and $\mathbf{r}'$ are associated with the same parent atom.
A practical implementation of this idealized model would be an infinite chain of helium atoms in the limit of infinitely large interatomic distances.

In order to capture the effect of unphysical delocalization, we first define a minimal basis set $\chi_A(\mathbf{r})$, where $A,B,C,\ldots$ denotes an atomic index. This basis set is designed such that it exactly captures the \emph{inter}-atomic contributions (i.e.~those which are off-block-diagonal in the density matrix), but is inexact in its description of \emph{intra}-atomic contributions (i.e. those which are block-diagonal in the density matrix). In other words, the total energy is stationary with respect to orbital rotations between atoms, but not stationary with regards to intra-atomic orbital rotations if it were extended by a complete basis set.  
Since in a minimal basis set, $\mathbf{P} = \mathbf{S}^{-1}$ (\cref{eq:P_min}) 
and off-block-diagonal contributions are zero by construction,
the conditions
\begin{equation}\label{eq:metric}
\mathbf{P} = \mathbf{S}^{-1} = \mathbf{1} = \mathbf{S}
\end{equation}
are necessary (up to a unitary transformation) to fulfill the requirement of zero off-diagonal atom blocks.
This fact as well as the choice of a normalized basis set straightforwardly imply the following overlap relations:
\begin{equation}
  S_{\chi\chi}^{AB} = \int \mathrm{d} \mathbf{r} \chi_A(\mathbf{r}) \chi_B(\mathbf{r}) = \delta_{AB}
\end{equation}
 
Next, this basis set is augmented with a diffuse basis function $\Delta_A(\mathbf{r})$ on each atom. We construct this diffuse basis function such that it overlaps with both basis functions on each neighbouring atom, but not others:
\begin{align}\label{eq:nearest_neighbour_1}
  S_{\Delta\chi}^{AB} = \int \mathrm{d} \mathbf{r} \Delta_A(\mathbf{r}) \chi_B(\mathbf{r}) &\neq 0  \quad A = B \pm 1 \\
  S_{\Delta\Delta}^{AB} = \int \mathrm{d} \mathbf{r} \Delta_A(\mathbf{r}) \Delta_B(\mathbf{r}) &\neq 0 \quad A = B \pm 1  \label{eq:nearest_neighbour_2}
\end{align}
Introduction of $\Delta_A(\mathbf{r})$ thus improves the \emph{intra}-atomic description of the present electronic structure (simply by introducing new degrees of freedom), but may deteriorate its \emph{inter}-atomic description.
Note that in this simplified model the augmentation functions only overlap with direct neighbors and not second, third and farther neighbors 
as would be the case for Gaussian basis sets.
The aim of this approach is to show that just nearest neighbor overlap is already sufficient 
to explain the observed non-locality caused by diffuse basis sets 
and overlap over larger distances would, in practice, just add even more non-locality.
In addition, we require the augmentation function to be block-orthogonal to the initial minimal basis function
\begin{equation}\label{eq:S_AA_dc}
  S^{AA}_{\Delta\chi} = \int \mathrm{d} \mathbf{r} \Delta_A(\mathbf{r}) \chi_A(\mathbf{r}) = 0,
\end{equation}
which may always be achieved through block-orthogonalization.

Finally, we define a new minimal basis set 
$\chi'_{A}(\mathbf{r})$ 
as an atom-wise linear combination of the two above defined basis sets
\begin{equation}\label{eq:chi_2}
  \chi'_{A}(\mathbf{r}) = N\big[\chi_A(\mathbf{r}) + \omega \Delta_A (\mathbf{r})\big]
\end{equation}
with a normalization constant
\begin{equation}
  N = (1+\omega^2)^{-\frac{1}{2}}.
\end{equation}
The linear contraction coefficient $\omega$ can be treated as an additional MO coefficient. Assuming that $\chi_A(\mathbf{r})$ -- while not exact -- is sufficiently close to the exact minimal basis, $\omega$ is small and can thus be approximated in first order 
($\lim_{\omega \rightarrow 0}$)
by its negative SCF energy gradient\cite{pulay_ab_1969}
\begin{equation}\label{eq:scf_gradient}
  \omega \approx -\frac{\partial E}{\partial \omega} = -2 F^{AA}_{\Delta \chi}
\end{equation}
where 
\begin{equation}
  F^{AA}_{\Delta \chi} = \int \mathrm{d}\mathbf{r} \Delta_A (\mathbf{r}) \hat{F} \chi_A(\mathbf{r})
\end{equation}
denotes the mixed-basis Fock matrix elements and $\hat{F}$ the corresponding Fock-operator. 
The derivation of \cref{eq:scf_gradient} is given in Appendix~\ref{sec:pulay}. 
This procedure is equivalent to a single step of gradient descent without a scaling factor.
Under the assumption that $\chi_A(\mathbf{r})$ is inexact and due to the requirement that the SCF energy be differentiable with respect to the MO coefficients, there always exists a sufficiently small scaling factor where a lowering of the energy is guaranteed.
We can therefore quantify $\omega$ to be proportional to its first order energy gradient (i.e.~\cref{eq:scf_gradient}).
The main purpose of this approximation is to show how mixing in a diffuse function can, in principle,
lead to an artificially delocalized solution which is guaranteed to be lower in energy and thus variationally favored.

We will now construct the overlap matrix in the basis of $\chi'_A(\mathbf{r})$. The ansatz in \cref{eq:chi_2} together with the nearest-neighbor overlap conditions (\cref{eq:nearest_neighbour_1,eq:nearest_neighbour_2})
leads to the tridiagonal form 
\begin{equation}
  \mathbf{S} = \mathrm{trid}(t,1,t) \equiv
  \begin{pmatrix}
    \ddots & \vdots & \vdots & \vdots & \vdots & \vdots & \iddots \\
    \dots & t & 1 & t & 0 & 0 &\dots\\
    \dots & 0 & t & 1 & t & 0 &\dots\\
    \dots & 0 & 0 & t & 1 & t &\dots\\
    \iddots & \vdots & \vdots & \vdots & \vdots & \vdots & \ddots \\
  \end{pmatrix}
\end{equation}
where the off-diagonal element $t$ is easily derived as 
\begin{equation} 
  t = 4 N^2 (F^{AA}_{\Delta \chi} S^{AB}_{\Delta\chi} + (F^{AA}_{\Delta \chi})^2 S_{\chi\chi}^{AB})
\end{equation}
by inserting \cref{eq:scf_gradient} into the linear expansion of \cref{eq:chi_2}.
In the limit of a small gradient or, equivalently, small $\omega$, this reduces to 
\begin{equation} 
  \mathbf{S} = N^2 \mathrm{trid}(s,1+s^2,s),
\end{equation}
where $s$ has the following proportionality condition:
\begin{equation} \label{eq:s}
  s \propto F^{AA}_{\Delta \chi} S^{AB}_{\Delta\chi}
\end{equation}
The matrix inverse of this tridiagonal Toeplitz matrix is exactly known\cite{demko_decay_1984} as
\begin{equation} \label{eq:S_inv_tridiagonal}
  (\mathbf{S}^{-1})_{\mu\nu} = (-s)^{|\mu - \nu|}
\end{equation}
which decays exponentially with respect to the distance from the diagonal $|\mu - \nu|$.

\section{Understanding the curse of sparsity}\label{sec:understanding}
In combination, \cref{eq:s,eq:S_inv_tridiagonal} explain why even the quite local basis set $\chi'$, which only features nearest-neighbor overlap,
still leads to exponential decay in $\mathbf{S}^{-1}$ and 
consequently also $\mathbf{P}$ 
because the augmented basis is still minimal and therefore
$\mathbf{P} = \mathbf{S}^{-1}$ (\cref{eq:P_min}) still holds.
This means, we were able to reconstruct the \enquote{curse of sparsity} 
with this simple model system of non-interacting helium atoms.
This provides some useful insights into the nature of the problem:
\begin{enumerate}
  \item Even a quite local basis set (only nearest-neighbor overlap) can still lead to exponential decay in $\mathbf{S}^{-1}$ and $\mathbf{P}$. 
    In particular, $\mathbf{S}^{-1}$ can be substantially more delocalized than $\mathbf{S}$.
  \item The rate of the exponential decay is proportional to the nearest neighbor overlap $S^{AB}_{\Delta\chi}$
and the \emph{intra}-atomic Fock-matrix element $F^{AA}_{\Delta \chi}$, which (through \cref{eq:scf_gradient})
may be regarded as how well the augmentation function $\Delta(\mathbf{r})$ improves the description of the imperfect un-augmented basis set
$\chi(\mathbf{r})$.
  Thus, the \enquote{curse of sparsity} is most severe for small, yet diffuse basis sets such as aug-cc-pVDZ, explaining the results of \cref{fig:He10_dna_P_rs}.
  \item The exponential decay is oscillatory due to the alternating nature of $(-s)^x$, explaining the oscillations in \cref{fig:He10_dna_P_rs}.
  \item The exponential decay rate can, in principle, be made arbitrarily small by adding very diffuse functions without improving 
    \enquote{local} (e.g., intra-atomic) basis set completeness at the same pace.
    This artificial non-locality is thus a pure basis artifact and is independent of the \enquote{true} (i.e. basis set converged)
    electronic sparsity of the system.
\end{enumerate}

\section{Possible solution: CABS-singles correction}\label{sec:cabs}
After characterizing the problem in depth, 
one naturally asks for a full resolution of the conundrum at best, 
or at least partial remedies at worst.
Conceptionally, there exist two fundamental approaches to a satisfactory solution:
Either,  the use of diffuse basis sets is avoided altogether 
and one then tries to resolve the accompanying accuracy problems, i.e., \enquote{blessing compact basis sets with accuracy}, 
or one somehow constrains the SCF solution to be sufficiently local, even for diffuse basis sets, 
i.e.~\enquote{lifting the curse of sparsity}.
One possible solution using the former approach of avoiding diffuse basis sets and then \textit{a posteriori} restoring the accuracy is discussed below:

A classical approach to reduce basis set errors from SCF calculation is in the form of a dual basis approach
where a smaller basis is used to converge to a stationary SCF solution and then a second, larger basis set is employed as a one-shot correction.\cite{liang_approaching_2004,steele_dual-basis_2006,steele_non-covalent_2009,steele_ab_2010,hegely_dual_2018,csoka_speeding_2021,adler_simple_2007,shaw_approaching_2017}
One particular approach is a CI-singles correction where a complementary auxiliary basis set (CABS) is introduced perturbatively:
\cite{adler_simple_2007, shaw_approaching_2017}
\begin{equation}
  E_s = \sum_{i\alpha} t_i^\alpha F_{i\alpha},
\end{equation}
where the singles amplitudes $t_i^\alpha$ are obtained by solving the linear equation
\begin{equation}
  F_{i\alpha} = \sum_k t_k^\alpha F_{ik} - \sum_\beta F_{\alpha \beta}t_i^\beta,
\end{equation}
and where $i,j,k$ denote occupied molecular orbitals (MOs) and $\alpha$, $\beta$ denote orthogonalized CABS basis functions.\cite{adler_simple_2007, shaw_approaching_2017}
This approach is typically combined with the F12-method for post-SCF correlation calculations,
but may just as well be applied to free-standing SCF calculations.

Thus, we decided to employ the CABS singles approach to improve the accuracy of the compact cc-pVXZ (X=D,T,Q,5,6) basis sets by 
utilizing the corresponding augmented aug-cc-pVXZ basis sets as the respective CABS basis sets.
This particular choice of basis sets combinations is quite computationally efficient since the augmented basis sets are true supersets of their corresponding compact counterparts,
such that the complementary basis sets contain only diffuse functions and no additional compact functions are added.
This reduces the computational cost of the necessary Fock-build in the combined basis sets substantially, 
which is further improved by using seminumerical integration (sn-LinK) 
and resolution-of-the-identity Coulomb (RI-J) which we recently implemented 
for these kind of mixed-basis Fock-builds.\cite{urban_highly_2022}
The so obtained results are presented in \cref{tab:s22}.

\begingroup
\squeezetable
\begin{table}[htbp]
  \caption{Basis set errors (Hartree-Fock) [\si{\kilo\cal\per\mol}] 
  referenced to aug-cc-pV6Z as root-mean-square-deviation (RMSD), maximal deviation (Max)
  and mean signed error (MSE) for S22 benchmark (non-covalently bound dimers).\cite{jurecka_benchmark_2006,grafova_comparative_2010}
  No counterpoise corrections are applied. 
  Timings are given in seconds for one full Hartree-Fock SCF calculation for an \ch{(AT)4}-DNA fragment (260~atoms).
  Additional computational details are provided in appendix \ref{sec:comp_details}.
  \label{tab:s22}
  }
  \begin{ruledtabular}
  \begin{tabular}{lrrrr}
    Basis set combination & \multicolumn{1}{c}{RMSD} & \multicolumn{1}{c}{Max} & \multicolumn{1}{c}{MSE} & Time [s]\\ 
    \colrule 
    cc-pVDZ                             & 2.074 & 4.014 & -1.735 &   126 \\
    cc-pVTZ                             & 0.717 & 1.602 & -0.622 &   457 \\
    cc-pVQZ                             & 0.288 & 0.608 & -0.255 &  1571 \\
    cc-pV5Z                             & 0.057 & 0.111 & -0.048 &  6079 \\
    cc-pV6Z                             & 0.015 & 0.028 & -0.012 & 14279 \\
    aug-cc-pVDZ                         & 0.953 & 1.948 & -0.805 &   693 \\
    aug-cc-pVTZ                         & 0.228 & 0.493 & -0.195 &  2105 \\
    aug-cc-pVQZ                         & 0.057 & 0.116 & -0.050 &  6434 \\
    aug-cc-pV5Z                         & 0.005 & 0.011 & -0.004 & 22362 \\
    aug-cc-pV6Z                         &     - &     - &      - & 54036 \\
    cc-pVDZ$\rightarrow$aug-cc-pVDZ     & 1.343 & 2.879 & -1.205 &   201 \\
    cc-pVTZ$\rightarrow$aug-cc-pVTZ     & 0.385 & 0.848 & -0.343 &   680 \\
    cc-pVQZ$\rightarrow$aug-cc-pVQZ     & 0.119 & 0.247 & -0.107 &  2357\\ 
    cc-pV5Z$\rightarrow$aug-cc-pV5Z     & 0.017 & 0.034 & -0.015 &  8355\\ 
    cc-pV6Z$\rightarrow$aug-cc-pV6Z     & 0.003 & 0.007 & -0.003 & 19996\\ 
    cc-pV6Z$\rightarrow$aug-cc-pVTZ     & 0.003 & 0.007 & -0.003 & 17834\\
    prune-cc-pVQZ\footnotemark[1]       & 0.333 & 0.714 & -0.291 &   594 \\
    prune-cc-pV5Z\footnotemark[1]       & 0.080 & 0.163 & -0.068 &  1197 \\
    prune-cc-pV6Z\footnotemark[1]       & 0.030 & 0.057 & -0.025 &  2093 \\
    prune-cc-pVQZ\footnotemark[1]$\rightarrow$aug-cc-pVTZ
                                        & 0.156 & 0.366 & -0.135 & 884 \\
    prune-cc-pV5Z\footnotemark[1]$\rightarrow$aug-cc-pVTZ
                                        & 0.036 & 0.070 & -0.033 & 1650 \\
    prune-cc-pV6Z\footnotemark[1]$\rightarrow$aug-cc-pVTZ
                                        & 0.021 & 0.055 & -0.017 & 2757 \\
    prune-cc-pV6Z\footnotemark[1]$\rightarrow$aug-cc-pVQZ 
                                        & 0.020 & 0.034 & -0.018 & 2992 \\
  \end{tabular}
  \end{ruledtabular}
  \footnotetext[1]{pruned to contain fewer high-lqn Gaussians as defined in appendix~\ref{sec:comp_details}.}
\end{table}
\endgroup

Analogously to the ASCDB results of \cref{tab:basisset_errors}, augmentation with diffuse functions reduces the basis set errors for the NCIs in the S22 test set\cite{jurecka_benchmark_2006, grafova_comparative_2010} very substantially, 
e.g.~from \SI{0.717}{\kilo\cal\per\mol} for cc-pVTZ to \SI{0.228}{\kilo\cal\per\mol} for aug-cc-pVTZ,
although this improvement comes with a significant increase in computational cost.
Moreover, the CABS-singles correction significantly reduces the error of the original 
unaugmented basis set, despite not quite reaching the accuracy of its fully augmented counterpart.
E.g.~cc-pVTZ$\rightarrow$aug-cc-pVTZ halves the error of cc-pVTZ from \SI{0.717}{\kilo\cal\per\mol} to \SI{0.385}{\kilo\cal\per\mol}
and only falls short slightly compared to the full aug-cc-pVTZ results of \SI{0.228}{\kilo\cal\per\mol},
while only modestly increasing the computational cost from \SI{457}{\second} to \SI{680}{\second} which is a substantial speedup over the \SI{2105}{\second} of the full aug-cc-pVTZ calculation.
Interestingly, the size of the CABS basis set has a surprisingly small impact and aug-cc-pVTZ might generally be sufficient as a CABS basis, 
e.g.~cc-pV6Z$\rightarrow$aug-cc-pVTZ and cc-pV6Z$\rightarrow$aug-cc-pV6Z provide virtually the same results.

In addition, we noticed that contributions from high l-quantum number (lqn) basis functions are insignificant for SCF energies 
and are instead expected to only contribute substantially to post-SCF correlation calculations.
Thus, we decided to construct a pruned version of the cc-pVXZ(X=Q,5,6) basis sets where most high lqn basis functions are removed 
as outlined in appendix~\ref{sec:comp_details}.
This approach appears to be very successful, especially in combination with the CABS-singles correction, 
e.g.~prune-cc-pV5Z$\rightarrow$aug-cc-pVTZ provides nearly basis set converged results with an RMSD of \SI{0.036}{\kilo\cal\per\mol} at a lower computational cost (\SI{1650}{\second}) than full aug-cc-pVTZ (\SI{2105}{\second}). 
Surprisingly, there seem to be diminishing returns in the application of the CABS-singles correction to prune-cc-pV6Z,
only improving from \SI{0.030}{\kilo\cal\per\mol} to \SI{0.020}{\kilo\cal\per\mol},
most likely because the radial components of prune-cc-pV6Z are already quite basis set complete 
and thus only a small improvement is achieved through augmentation with diffuse functions.

Finally, we repeated the full ASCDB benchmark of \cref{tab:basisset_errors} for the new prune-cc-pVXZ basis sets to
obtain a more comprehensive view of their accuracy and establish if the accuracy transfers from Hartree-Fock to DFT (i.e.~$\omega$-B97X-V).
The results are presented in \cref{tab:basisset_errors_2}.
\begin{table}[htbp]
  \caption{Same as \cref{tab:basisset_errors} but for prune-cc-pVXZ family of basis sets. 
  \label{tab:basisset_errors_2}
  }
  \begin{ruledtabular}
  \begin{tabular}{lrrrrrrrrr}
    Basis set & \multicolumn{1}{c}{RMSD (B)} & \multicolumn{1}{c}{RMSD (M+B)} & \multicolumn{1}{c}{NCI RMSD (B)} & \multicolumn{1}{c}{NCI RMSD (M+B)} & Time [s]\\
    \colrule
    Prune-cc-pVQZ  &  5.90 & 17.50 & 6.94 &  7.33 &  666\\
    Prune-cc-pV5Z  &  2.83 & 17.33 & 1.78 &  2.91 & 1310\\
    Prune-cc-pV6Z  &  0.81 & 16.76 & 0.56 &  2.45 & 2387\\
  \end{tabular}
  \end{ruledtabular}
\end{table}
Indeed, the prune-cc-pVXZ variants are only marginally less accurate than their respective unpruned counterparts 
for both the whole ASCDB benchmark as well as the NCI subset, while saving a very substantial amount of computational cost. 
In particular, prune-cc-pV5Z comes close to the overall accuracy of def2-TZVPPD 
-- \SI{17.33}{\kilo\cal\per\mol} vs.~\SI{16.40}{\kilo\cal\per\mol} for whole ASCDB 
and \SI{2.91}{\kilo\cal\per\mol} vs.~\SI{2.40}{\kilo\cal\per\mol} for NCIs --
at a slightly lower computational cost (\SI{1310}{\second} instead of \SI{1440}{\second})
and without the need to employ diffuese basis functions.
Thus, this approach of constructing less diffuse basis set with instead a larger amount of compact, low-lqn functions
appears to be another promising tool to solve or at least somewhat remedy the conundrum of diffuse basis sets.

\section{Conclusion and outlook}
\label{sec:conc}
In this work, the \enquote{conundrum of diffuse basis sets}, 
i.e., the fact that diffuse basis sets are essential for 
non-covalent interactions (\enquote{the blessing of accuracy}) but 
are detrimental to electronic sparsity (\enquote{the curse of sparsity}), 
was explored in-depth.
In particular, it was found, that diffuse basis sets 
result in a much slower exponential decay of the 1-PDM than the intrinsic electronic locality of the system would imply.
Moreover, this effect was found to be independent of representation and could also be reproduced in the real-space 1-PDM, 
hinting towards an ill-defined basis set limit of the asymptotic decay rate of the 1-PDM.
It was then discovered that the problem may be rooted 
in the non-locality of the contra-variant basis functions as characterized by their overlap matrix $\mathbf{S}^{-1}$, 
which, in practice, is much less local than its co-variant dual $\mathbf{S}$.

Next, employing the model system of an infinite chain of non-interacting helium atoms, 
this basis set artifact could be further characterized:
In particular, the exponential decay rate could be quantified to be proportional to 
the nearest neighbor overlap as well as the intra-atomic basis set incompleteness.
This insight may prove useful in the design of new basis sets, which need to be more compact and more locally complete in 
order to not suffer from this artificial non-locality.
One example of how to incorporate this insight into basis set design was given in this work in the form of the 
lqn-reduced prune-cc-pVXZ (X=Q,5,6) basis sets.
These provide quite accurate NCIs at a much lower computational cost as compared to their un-pruned counterparts 
and without employing diffuse basis functions.

Finally, the CABS singles correction was proposed as one possible resolution to the conundrum,
where only compact basis sets are used within the SCF procedure 
and diffuse basis functions are only introduced perturbatively in the form of a one-shot post-SCF CI-singles correction.

Since the \enquote{the curse of sparsity} essentially arises as an overfitting artifact associated with an optimization in a finite basis set,
a complete resolution of the problem seems possible in principle.
The ideal resolution would still allow for the use of diffuse basis sets 
and instead constrain the SCF solution to only contain local solutions, e.g., by removing non-local contra-variant basis functions from
the optimization manifold.
However, we could not come up with any algorithm that could successfully identify and remove non-local functions without sacrificing too much accuracy.
Thus, there is still significant room for further development in that area, especially considering that 
the \enquote{the curse of sparsity} is arguably one of the largest factors
holding back the application of linear-scaling methods to practical problems.

\begin{acknowledgments}
  The authors thank Dr.~J.~Kussmann (LMU Munich) for providing a development version of the FermiONs++ program package.
  HL acknowledges financial support by Deutsche Forschungsgemeinschaft (DFG) through the Walter Benjamin Program (no.~529879166).
  Work at Berkeley was also supported by the U.S. Department of Energy, Office of Science, Office of Advanced Scientific Computing, and Office of Basic Energy Sciences, via the Scientific Discovery through Advanced Computing (SciDAC) program.
\end{acknowledgments}

\section*{Supplementary  Material}
The supplementary material contains all individual data points for the ASCDB the S22 benchmark set as well as its extension 
to the S22x5 benchmark set\cite{grafova_comparative_2010} where NCI distances are scaled from the minimum distance.
In addition, the basis set files of the new prune-cc-pVXZ are also provided.

\section*{Data Availability}
The data that support the findings of this study are available within this article and its supplementary material.

\section*{Author Declarations}
\subsection*{Conflicts of interest}
The authors have no conflicts of interests to disclose.
\subsection*{Author Contributions}
\textbf{Henryk Laqua:} 
Conceptualization (equal); 
Data curation (lead); 
Formal analysis (equal); 
Methodology (equal);
Validation (equal); 
Software (lead); 
Visualization (lead); 
Funding acquisition (equal);
Writing – original draft (lead).
\textbf{Linus Bjarne Dittmer:}
Conceptualization (supporting); 
Formal analysis (supporting); 
Methodology (supporting);
Validation (supporting); 
Writing – review \& editing (equal).
\textbf{Martin Head-Gordon:} 
Conceptualization (equal); 
Formal analysis (equal);
Methodology (equal);
Validation (equal);
Project administration (lead);
Supervision (lead);
Funding acquisition (equal);
Writing – review \& editing (equal).

\appendix
\section{Computational Details}
\label{sec:comp_details}
All calculations were performed with the FermiONs++ program package\cite{kussmann_pre-selective_2013,kussmann_preselective_2015}
using the resolution-of-the-identity Coulomb (RI-J) approximation in combination with the seminumerical exact-exchange method sn-LinK.\cite{weigend_fully_2002,neese_improvement_2003,neese_efficient_2009,laqua_efficient_2018,laqua_highly_2020,laqua_accelerating_2021,kussmann_highly_2021,urban_highly_2022}

For each AO basis set, the corresponding RI-J basis sets were employed, e.g.~def2-TZVPP-RI-JK in combination with def2-TZVPPD or 
cc-pVTZ-JKFit in combination with (aug-)cc-pVTZ. \cite{weigend_fully_2002,weigend_hartree-fock_2007,gulde_error-balanced_2012}
Since no cc-pV6Z-JKFit basis set exists yet, the cc-pV5Z-JKFit basis sets were employed instead in these situations.
The impact of the Coulomb-fitting basis set is, however, insignificant in the context of this work and we particularly 
checked that the observed non-locality artifacts are not at all related to the RI-J approximation.

Similarly, the \enquote{gm5} integration grid\cite{laqua_improved_2018} employed in the seminumerical exact-exchange method
and the semi-local density functional approximation are also sufficiently accurate to not introduce significant errors within the context of this work.
Due to technical limitations, all basis functions with a $\text{lqn}>5$, i.e. I- and J- functions were removed from all basis sets.
We also expect virtually no impact on any of the results shown in this work from this removal.
Moreover, since (aug-)cc-pV6Z is only published for elements up to \ch{Ar}, (aug-)cc-pV5Z was utilized instead for all other elements. 
In addition, we had to employ the def2- basis sets and their respective effective core potentials (ECPs) for ruthenium
within all (aug)-cc-pVXZ basis set calculations, since no basis set definition for {Ru} was available.

The lqn reduced basis sets prune-cc-pVXZ (X=Q,5,6) utilize the radial functions of the parent (Q,5,6)-Z basis set for the minimal lqns, i.e.~s-functions for 
hydrogen and s- and p-functions for main-group elements and then employ one level lower for each higher lqn, e.g.~d-functions of cc-pV5Z and f-functions of cc-pVQZ for main-group elements in prune-cc-pV6Z.
The basis set files are provided as supplemental material.
Finally, all timings were performed on a dual-socket AMD EPYC-9334 (2$\times$32 cores@\SI{2.7}{\giga\hertz}) server node.

\section{The first order energy gradient\label{sec:pulay}}
Applying Pulay's SCF Gradient formula\cite{pulay_ab_1969} 
\begin{equation}
  E^x = \sum_{\mu\nu}P_{\mu\nu} (h_{\mu\nu}^x + \frac{1}{2} \sum_{\lambda\sigma} P_{\lambda\sigma}(\mu\nu||\lambda\sigma)^x)
  - \sum_{\mu\nu} W_{\mu\nu} S_{\mu\nu}^x
\end{equation}
to $E^x = \frac{\partial E}{\partial \omega_A}$
and collecting all terms results in 
\begin{equation}\label{eq:pulay}
  \frac{\partial E}{\partial \omega_A} = 2\sum_B (P^{AB}_{\chi\chi} F^{AB}_{\Delta\chi} - W^{AB}_{\chi\chi}S^{AB}_{\Delta\chi}),
\end{equation}
where $\mathbf{W}$ denotes the energy-weighted 1-PDM
\begin{equation}\label{eq:W}
  W_{\mu\nu} = \sum_i \varepsilon_i C_{\mu i} C_{\nu i} = (\mathbf{C} \mathbf{\epsilon} \mathbf{C}^T)_{\mu\nu}
\end{equation}
with molecular orbital (MO) coefficients $\mathbf{C}$ and corresponding orbital energies $\varepsilon$.

Due to the translational symmetry of the helium chain (all atoms are symmetry equivalent),
all eigenvalues $\varepsilon_i$ are identical, allowoing the matrix of Eigenvalues $\mathbf{\epsilon}$ 
to be decomposed as
\begin{equation}\label{eq:epsilon}
  \mathbf{\epsilon} = \varepsilon \mathbf{1},
\end{equation}
where $\varepsilon$ is one global scalar.
Inserting \cref{eq:epsilon} into the definition of $\mathbf{W}$ (\cref{eq:W} leads to
\begin{equation}
  \mathbf{W} = \mathbf{C} \mathbf{\epsilon} \mathbf{C}^T = \varepsilon\mathbf{C} \mathbf{1} \mathbf{C}^T = \varepsilon\mathbf{C} \mathbf{C}^T = \varepsilon\mathbf{P} = \varepsilon\mathbf{1},
\end{equation}
where we utilized the definition of the 1-PDM 
\begin{equation}
  \mathbf{P} = \mathbf{C}_\text{occ}\mathbf{C}^T_\text{occ} = \mathbf{C}\mathbf{C}^T
\end{equation}
and the fact that the basis is still minimal, i.e.~there is no virtual space ($\mathbf{C}_\text{occ} = \mathbf{C}$).
Finally, the identities
\begin{align}
  P^{AB}_{\chi\chi} &= 0\\
  S^{AA}_{\chi\Delta} &= 0\\
  S^{AB}_{\chi\chi} &= 0 
\end{align}
simplify \cref{eq:pulay} to yield \cref{eq:scf_gradient}.

\bibliography{lib}
\end{document}